\documentclass[aps,prapplied,twocolumn,superscriptaddress,floatfix]{revtex4-2}

\usepackage[utf8]{inputenc}

\usepackage{physics}
\usepackage{amsmath}
\usepackage{amssymb}
\usepackage{graphicx}
\usepackage{bm}

\usepackage{hyperref}
\hypersetup{
    colorlinks=true,allcolors=blue,
}

\newcommand{\Eodd}{E_{\rm odd}}
\newcommand{\dstar}{\delta_\star}
\newcommand{\phiext}{\phi_{\rm ext}}

\begin{document}

\title{Noise-Protected Superconducting Qudits from Higher-Harmonic Josephson Junctions}

\author{Hamed Vakili}
\affiliation{Department of Physics and Astronomy and Nebraska Center for Materials and Nanoscience, University of Nebraska, Lincoln, Nebraska 68588, USA}

\date{November 11, 2026}

\begin{abstract}
Qudits, which generalize qubits to $d$ levels, store more information per circuit element than qubits. As with qubits, however, their sensitivity to noise must be overcome before they can be used for fault-tolerant quantum computing. Here we show that, by exploiting the second harmonic of the current--phase relation, a superconducting loop of three Josephson junctions can host qutrits strongly protected against relaxation, while dephasing limits their coherence. When the second harmonic is positive, a simple loop with a small asymmetry between the junctions provides a tunable, symmetry-protected qutrit. When it is negative, a protected qutrit can be realized by adding a small symmetry-breaking term, such as a weak Josephson diode effect. The two designs involve different trade-offs: the first has an exact charge selection rule, whereas the second is nearly insensitive to charge noise and its quasiparticle errors can be corrected. The preferable choice therefore depends on the materials used. Both require a sizable second harmonic to begin with, and we identify altermagnetic Josephson junctions as a candidate platform. Finally, microscopic calculations show that a gate voltage selects the sign of the second harmonic in these junctions, although gate-voltage noise then becomes the leading source of dephasing. These findings establish the sign of the second harmonic as a design principle for protected superconducting qudits.
\end{abstract}

\maketitle

Superconducting circuits are among the leading candidates for implementing qubits, with Josephson junctions providing the anharmonicity needed to address them. They combine this strong nonlinearity with lithographic scalability and fast electrical control \cite{Koch2007,Devoret2013superconducting,Blais2021}. However, their coherence is limited by dielectric loss, quasiparticles, and charge and flux fluctuations. One route toward fault tolerance is to build noise suppression into the circuit Hamiltonian, so that the computational subspace couples weakly to the dominant noise operators while remaining addressable \cite{Kitaev2006,Brooks2013zeropi,Manucharyan2009fluxonium,Gyenis2021}. An example is a Josephson element whose current--phase relation (CPR) carries a sizable second harmonic, which transfers pairs of Cooper pairs coherently \cite{Smith2020,Doucot2012}. Such elements are obtained by engineered interference \cite{Gladchenko2009,Bell2014,Larsen2020,Roverch2026,Zhurbina2026}, or natively in twisted $d$-wave junctions near $45^\circ$ \cite{Sigrist1995,Tanaka1997,Lofwander2001,Can2021,Zhao2023,Tummuru2022,Volkov2024} and in altermagnetic junctions near a $0$--$\pi$ transition \cite{Ouassou2023,BeenakkerVakhtel2023,Lu2024phi,Sun2025}. Devices built on them have so far encoded \emph{qubits} \cite{Smith2020,Larsen2020,Roverch2026,Messelot2026,Brosco2024,VosoughiNia2026,Tjernshaugen2026,Guo2026altermon,Pan2026am}, and their protection has been argued from charge and quasiparticle matrix elements between the computational states. Extending protection to \emph{qudits} increases the information encoded per circuit element and enriches the routes to error correction \cite{Neeley2009,ChirolliMoore2021,Fraxonium2026}. Here we show that a three-junction loop of such elements hosts protected qutrits ($d=3$ qudits) in both positive and negative sign classes of the second CPR harmonic, with the first harmonic taken positive. In the positive class, the junctions need only a sizable second harmonic and a small asymmetry between them. In the negative class, the junctions must also carry small, unequal energy terms odd in the phase, as in a Josephson diode. These can be supplied by intrinsically nonreciprocal junctions \cite{Volkov2024,Vakili2024jde,Cheng2024amdiode,Sahoo2026amdiode} or auxiliary side junctions. We find that dephasing rather than relaxation limits both classes, and that the sign class decides whether quasiparticle errors can be corrected.

We take the harmonics of each junction as given. A junction with phase $\delta$ across it has the energy $U(\delta)=-\sum_{m\ge1}E_m\cos(m\delta)$ with harmonic vector $(E_1,E_2,\dots)$, and its sign class is set by the ratio
\begin{equation}
\eta \;\equiv\; \frac{E_2}{|E_1|} \;=\; \frac{I_2}{2|I_1|},
\label{eq:eta_def}
\end{equation}
with $I_m$ the CPR harmonics, $\eta>0$ the positive class and $\eta<0$ the negative one. The first harmonic enters through its magnitude because shifting $\delta$ by $\pi$ changes the sign of $E_1$ but not $E_2$. We measure energies in units of $|E_2|$.

\begin{figure}[t]
  \centering
  \includegraphics[width=\columnwidth]{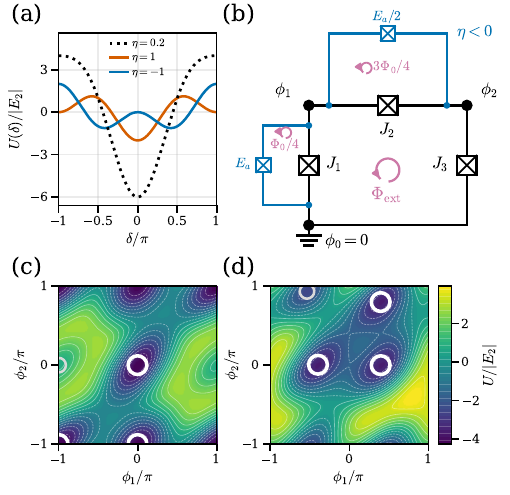}
  \caption{The two sign classes. (a) Junction energy $U(\delta)/|E_2|$ for three values of $\eta$; the dotted curve at $\eta=0.2$ lies below the threshold $|\eta|=1/4$. (b) The circuit. Three higher-harmonic elements $J_j$, each shunted by a capacitor (not drawn), form a loop threaded by $\Phi_{\rm ext}$; this is the whole $\eta>0$ circuit. For $\eta<0$, side junctions $E_a$ and $E_a/2$ (blue) outside the loop, biased by the same field, close sub-loops holding $\Phi_0/4$ and $3\Phi_0/4$ and so add $+E_a\sin\delta_1$ and $-(E_a/2)\sin\delta_2$, the odd term of Eq.~\eqref{eq:odd_split}. (c) Potential of the loop for $\eta=+0.8$ at $\phiext=\pi$: the three $(\pi,0,0)$ permutations (white) are degenerate, and $(\pi,\pi,\pi)$ (gray) lies $4E_1$ higher. (d) The same for $\eta=-0.8$ at $\phiext=\dstar$, before the odd term: the three $(\dstar,\dstar,-\dstar)$ permutations (white) are degenerate, and a fourth well (gray) lies higher.}
  \label{fig:circuit}
\end{figure}

\emph{Circuit.}---The circuit is a loop of three branches, each a Josephson element shunted by a capacitor of charging energy $E_C$ [Fig.~\ref{fig:circuit}(b)]. The phases across the elements are $\delta_1=\phi_1$, $\delta_2=\phi_2-\phi_1$ and $\delta_3=\phiext-\phi_2$, with node phases $\phi_{1,2}$ and reduced flux $\phiext=2\pi\Phi_{\rm ext}/\Phi_0$ \cite{Orlando1999,DevoretLesHouches1995,VoolDevoret2017}. The Hamiltonian is
\begin{equation}
\hat H=4(\hat{\mathbf n}-\mathbf n_g)^{T}\mathbf{E}_C(\hat{\mathbf n}-\mathbf n_g)
+\sum_{j=1}^{3}U_j(\hat\delta_j),
\label{eq:H_final}
\end{equation}
where $\mathbf E_C=(e^2/2)\mathbf C^{-1}$ is the charging-energy matrix, $\mathbf C$ the capacitance matrix, $\hat{\mathbf n}$ are the node charges and $\mathbf n_g$ the offset charges. The eigenvalues are $\mathcal E_k$, and the three lowest states are the computational states. With two harmonics, a junction has two wells only when $|\eta|>1/4$ [Fig.~\ref{fig:circuit}(a)]. In the positive class, $\eta>1/4$, the wells lie at $\delta=0$ and $\pi$ and differ in energy by $2E_1$. In the negative class, $\eta<-1/4$, they lie at $\pm\dstar$ with $\cos\dstar=1/(4|\eta|)$. The loop then fixes the operating flux, at which its lowest wells are degenerate [Fig.~\ref{fig:circuit}(c,d)]: $\phiext=\pi$, half a flux quantum, for $\eta>0$, and $\phiext=\dstar$ for an $\eta<0$ qutrit or $\pi-\dstar$ for a four-level qudit \cite{SM}.

\emph{Three exact symmetries.}---\emph{S1, branch reflection:} when the CPR is odd, $U(\delta)$ is even, so the $\eta<0$ wells $\pm\dstar$ are degenerate for any harmonic content. \emph{S2, branch permutation:} identical branches make the computational states a $C_3$ multiplet at every flux, with equal loop currents. \emph{S3, loop reflection at half flux:} at $\phiext=\pi$ the reflection $(\phi_1,\phi_2)\to(-\phi_1,-\phi_2)$ reverses every $\delta_j$ and leaves Eq.~\eqref{eq:H_final} unchanged at $\mathbf n_g=0$, while the node charges change sign. For $\eta>0$ the wells lie at $\delta_j=0$ or $\pi$, the fixed points of this reflection. Each $(\pi,0,0)$ configuration then maps to itself, the three computational states have the same parity, and
\begin{equation}
\bra{i}\hat n_\alpha\ket{j}=0
\qquad\text{identically, } i,j\in\{0,1,2\},
\label{eq:selection_rule}
\end{equation}
for any $E_C$, harmonic vector or branch asymmetry. The projected charge norm $\mathcal N_{\mathcal Q}=(\sum_\alpha\|P\hat n_\alpha P\|^2)^{1/2}$, where $P$ projects on the computational subspace, stays at the numerical floor for every $|E_2|/E_C$ computed \cite{SM}. Offset charge relaxes the rule linearly in $n_g$. The loop current is odd under S3, so its expectation value vanishes and the computational states have no first-order flux sensitivity.

\emph{The odd term.}---The $\eta<0$ class has no such rule. Its three lowest wells at $\phiext=\dstar$ differ only in which branch sits at $-\dstar$, and S1 keeps them degenerate. A perturbation that keeps the branch potentials even (higher harmonics, branch asymmetry, unequal capacitors) therefore cannot split them at first order. What splits them is a term \emph{odd} in $\delta$. Adding $A_j\sin\delta_j$ shifts well $k$, in which branch $k$ sits at $-\dstar$, by
\begin{equation}
\Delta\mathcal E_k=\sin\dstar\Bigl(\sum_jA_j-2A_k\Bigr),
\label{eq:odd_split}
\end{equation}
which splits the wells only if the $A_j$ differ. The choice $A=(E_a,-E_a/2,0)$ keeps the two transitions distinct. With $E_a=0.05$ a $\sin\delta$ term splits the wells by about $0.1$ and a $\sin3\delta$ term by a third of that after zero-point averaging, while $\cos\delta$ or $\cos3\delta$ terms act only at second order, about a hundred times weaker \cite{SM}. Equivalently, the first harmonic is phase-shifted relative to the second by a few degrees, which is the condition for a Josephson diode \cite{Souto2022,Baumgartner2022,Coraiola2024,Vakili2024jde}, here of about one percent efficiency. Ordinary junctions supply the odd term. A small junction of Josephson energy $E_a$ across branch $j$, closing a sub-loop that holds $\Phi_0/4$ or $3\Phi_0/4$, adds $-E_a\cos(\delta_j\pm\pi/2)=\pm E_a\sin\delta_j$ exactly [Fig.~\ref{fig:circuit}(b)], as in the asymmetrically threaded SQUID of cat qubits \cite{Lescanne2020}. At that bias the sine amplitude is stationary, and each sub-loop is about as sensitive to flux as the main loop. One coil can bias all three loops when their areas are in the ratio of about $1:3:5$, the main loop holding one extra flux quantum. Percent-level area errors then move the levels by about $10^{-3}$ of the splitting \cite{SM}.

\emph{Figures of merit.}---Both classes meet the matrix-element criteria used for single-junction qubits \cite{Pan2026am,Brosco2024}, but these criteria say little here. The element $|\bra{0}\hat n\ket{0}-\bra{1}\hat n\ket{1}|^2$ vanishes at $n_g=0$ for every design, a transmon included, and the quasiparticle criterion assumes the thermal quasiparticle density, which is negligible beside the measured fraction $x_{\rm qp}\sim10^{-9}$--$10^{-6}$ \cite{Serniak2018}. We therefore evaluate every noise channel. Dielectric loss with quality factor $Q$ relaxes computational state $i$ at the rate
\begin{equation}
\Gamma_1=\sum_{j=1}^{3}\sum_{k<i}\frac{|\bra{k}\hbar\dot{\hat\delta}_j\ket{i}|^2}{4E_{Cj}\,\hbar Q},
\label{eq:gamma1}
\end{equation}
where $E_{Cj}$ are the branch charging energies and $\hbar\dot{\hat\phi}_\alpha=8\sum_\beta E_{C,\alpha\beta}(\hat n_\beta-n_{g\beta})$. The rate is thus set by the charge matrix elements $|\bra{i}\hat{\mathbf n}\ket{j}|=(\sum_\alpha|\bra{i}\hat n_\alpha\ket{j}|^2)^{1/2}$ \cite{SM}. Quasistatic noise of amplitude $\sigma$ dephases a qutrit transition at the rate $\hbar\Gamma_\varphi=s\sigma$, where $s$ is the sensitivity of its energy \cite{Ithier2005,Koch2007}. For charge noise we take $s=W$, the charge dispersion \cite{Koch2007}, the peak-to-peak variation of a qutrit transition energy with $\mathbf n_g$, and $\sigma_{n_g}=10^{-3}$. For critical-current noise we take $s=q$, the largest slope of a qutrit splitting per unit fractional change $u$ of each junction's harmonic vector, the junctions fluctuating independently, and $\sigma_u=10^{-6}$, its level in tunnel junctions \cite{VanHarlingen2004}. A quasiparticle tunneling across junction $j$ is an event that acts through $M_{fi}=\bra{f'}\sin(\hat\delta_j/2)\ket{i}$, with $\ket{f'}$ in the charge-parity sector the event switches to. Its rate is $|M_{fi}|^2S_{\rm qp}(\omega_{if})$, where $S_{\rm qp}\propto x_{\rm qp}$ is the quasiparticle spectral density \cite{Catelani2011}. We take $x_{\rm qp}=10^{-8}$, a superconducting gap $\Delta=5$, $k_BT=0.04$ and a junction energy $E_J=1$ in the prefactor of $S_{\rm qp}$. Because $S_{\rm qp}$ diverges for events that exchange no energy, we cut it off at frequencies between $10^{-6}$ and $10^{-3}$ of $|E_2|/h$ and show the spread as bands \cite{SM}. Such events dephase states $k$ and $l$ at the rate $\tfrac12|M_{kk}-M_{ll}|^2S_{\rm qp}$, summed over junctions, or $\tfrac12(|M_{kk}|-|M_{ll}|)^2S_{\rm qp}$ when each event is detected and its sign flip corrected, which we call tracked.

\begin{figure}[t]
  \centering
  \includegraphics[width=\columnwidth]{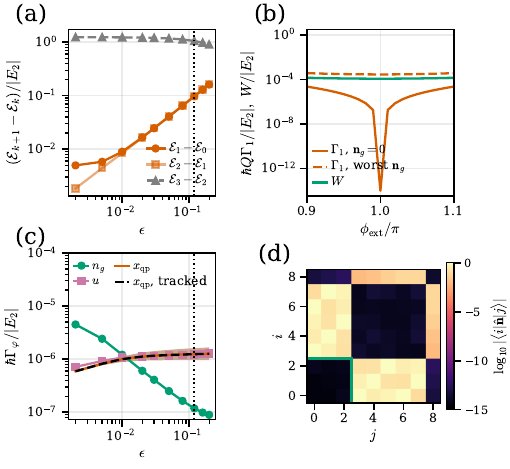}
  \caption{The $\eta>0$ qutrit: the harmonic vector of Eq.~\eqref{eq:am_vector} on branches scaled by $1+\epsilon$, $1$ and $1-\epsilon$, at $|E_2|/E_C=10$. (a) Level spacings versus $\epsilon$ at $\phiext=\pi$; $\mathcal E_3-\mathcal E_2$ is the gap above the computational states. (b) Relaxation rate of the faster-decaying state, Eq.~\eqref{eq:gamma1}, at $\mathbf n_g=0$ and maximized over $\mathbf n_g$, and the charge dispersion $W$, versus flux at $\epsilon=0.12$; at $\phiext=\pi$ the $\mathbf n_g=0$ rate vanishes to machine precision and is drawn at the floor. (c) Dephasing rates at $\phiext=\pi$ from charge noise ($\sigma_{n_g}=10^{-3}$), critical-current noise ($\sigma_u=10^{-6}$) and quasiparticles ($x_{\rm qp}=10^{-8}$; shaded bands show the spread over the cutoff of $S_{\rm qp}$); tracked and untracked rates coincide. (d) $|\bra{i}\hat{\mathbf n}\ket{j}|$ for the nine lowest states at $\epsilon=0.12$, $\phiext=\pi$, $\mathbf n_g=0$: the computational block (outlined) is empty. Dotted lines: operating point.}
  \label{fig:pos}
\end{figure}

\emph{The $\eta>0$ qutrit.}---With identical branches the computational states form a $C_3$ multiplet, split only by tunneling. An asymmetry $\epsilon$ that scales the three harmonic vectors by $1+\epsilon$, $1$ and $1-\epsilon$ splits them linearly [Fig.~\ref{fig:pos}(a)]. It does not break S3, which maps each branch onto itself. We take the altermagnet vector of Eq.~\eqref{eq:am_vector} at $|E_2|/E_C=10$ and $\epsilon=0.12$, which puts the computational levels $0.1$ apart beneath a gap of about $1$. Relaxation vanishes at $\phiext=\pi$ and $\mathbf n_g=0$ and returns away from either point [Fig.~\ref{fig:pos}(b)]. Even at the worst offset, $\hbar Q\Gamma_1$ is a few times $10^{-4}$. The dephasing rate $\hbar\Gamma_\varphi$ depends on what produces the splitting [Fig.~\ref{fig:pos}(c)]. The splittings are differences of junction energies, so critical-current noise enters with $q\approx1$ for every $\epsilon$, and quasiparticles dephase at a comparable rate. Charge noise is ten times weaker at the operating point and fades as $\epsilon$ localizes the wells. The computational block of the charge operator is empty [Fig.~\ref{fig:pos}(d)], so the qutrit is driven through a plasmon level above the computational states, shared by all three. This intermediate level survives the asymmetry only in shallow wells, which is why we choose $|E_2|/E_C=10$. A shaped two-tone pulse through it transfers $\ket{0}\to\ket{2}$ fast and with high fidelity \cite{Caneva2011,Motzoi2009,SM}.

\begin{figure}[t]
  \centering
  \includegraphics[width=\columnwidth]{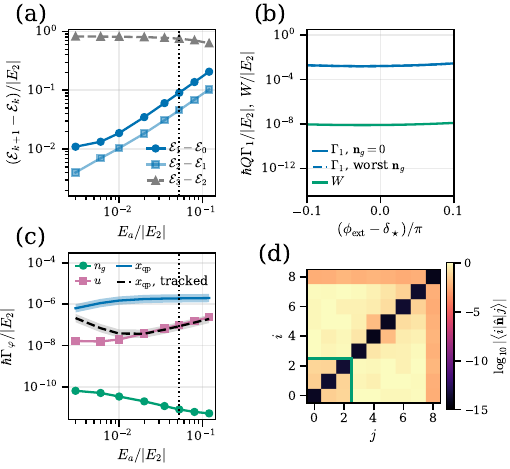}
  \caption{The $\eta<0$ qutrit: two harmonics with $|\eta|=0.8$ at $|E_2|/E_C=20$ and $\phiext=\dstar$, with side junctions supplying the odd term $A=(E_a,-E_a/2,0)$. (a) Level spacings versus $E_a$. (b) As Fig.~\ref{fig:pos}(b), versus the main-loop flux at $E_a=0.05$, sub-loop biases fixed; the two rates coincide. (c) As Fig.~\ref{fig:pos}(c), versus $E_a$; parity tracking lowers the quasiparticle rate. (d) As Fig.~\ref{fig:pos}(d), at $E_a=0.05$: the computational block carries tunneling-suppressed elements.}
  \label{fig:neg}
\end{figure}

\emph{The $\eta<0$ qutrit.}---With $A=(E_a,-E_a/2,0)$ the level spacings approach $2E_a\sin\dstar$ and $E_a\sin\dstar$ once $E_a$ exceeds the tunneling splitting [Fig.~\ref{fig:neg}(a)]. We take $|\eta|=0.8$, $|E_2|/E_C=20$ and $E_a=0.05$, which puts the computational levels $0.09$ and $0.05$ apart beneath a gap of $0.75$. No symmetry forbids relaxation here. The charge elements between computational states are suppressed only by tunneling between deep wells, so $\Gamma_1$ is finite and insensitive to flux and offset charge [Fig.~\ref{fig:neg}(b)], with $\hbar Q\Gamma_1\approx2\times10^{-3}$. These elements make the qutrit directly drivable [Fig.~\ref{fig:neg}(d)]. Flat-top $\pi$ pulses of a charge drive reach high fidelity within a few tens of periods of the qutrit transitions \cite{SM}. The charge dispersion is negligible: $W\approx10^{-8}$, four orders of magnitude below the $\eta>0$ device. The reason is topological. Offset charge enters only through tunneling paths that wind a node phase by $2\pi$. Neighboring $\eta<0$ wells are joined through the barrier at $\delta_j=0$ without winding, while winding must cross $\delta_j=\pi$, where the barrier is more than three times higher. Critical-current noise of the main junctions shifts all three wells together, by S1. It therefore acts through the side junctions and grows with $E_a$ [Fig.~\ref{fig:neg}(c)].

\emph{Quasiparticles.}---Each quasiparticle tunneling event acts through the computational block of $M$. If the moduli of its diagonal elements are unequal, the event measures the state. If the block is proportional to a unitary, the event causes a known error, which can be corrected once the event is detected. The sign class decides which case applies. For $\eta>0$ the wells sit at the fixed points $0$ and $\pi$, where $|\sin(\delta/2)|$ is $0$ or $1$. Each event therefore measures which branch sits at $\pi$, an error that cannot be corrected [Fig.~\ref{fig:pos}(c)]. The property that makes the selection rule exact, wells at the fixed points, thus makes every event a measurement. For $\eta<0$ the mirror wells share $|\sin(\dstar/2)|$, and an event across junction $j$ only flips the sign of the state in which branch $j$ sits at $-\dstar$. The charge parities of the two nodes identify $j$. Tracking these parities \cite{Serniak2019} with a separate detector reveals each event, so the flip can be corrected and the dephasing rate falls by more than an order of magnitude [Fig.~\ref{fig:neg}(c)]. Side junctions reopen a weak measurement channel, because across a junction biased at $\pm\pi/2$ the mirror wells give unequal $|M_{ii}|$. It grows with $E_a$, while at small $E_a$ the wells are not yet localized, so the tracked rate is lowest near $E_a\approx0.01$--$0.02$. The same model classifies single-junction designs: each event measures the $2\varphi$ qubit of Ref.~\cite{Pan2026am}, which stores its states in the $0$ and $\pi$ wells, but only flips a sign in the mirror-well flowermon \cite{Brosco2024}.

\begin{figure}[t]
  \centering
  \includegraphics[width=\columnwidth]{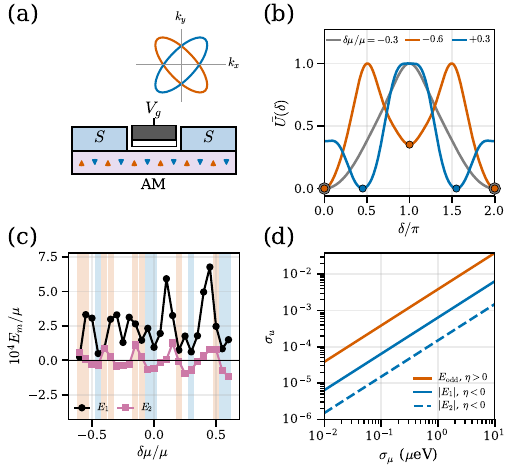}
  \caption{The altermagnetic platform. (a) A gate voltage $V_g$ sets the chemical potential $\mu+\delta\mu$ of the altermagnet (AM) between superconducting leads ($S$); its two spin species have Fermi surfaces elongated along orthogonal axes. (b) Junction energy $\bar U$, normalized to its range, at three gate voltages; circles mark the wells. (c) First and second harmonics per transverse channel versus $\delta\mu/\mu$, gauged to $E_1\ge0$; shading marks the $\eta>0$ (orange) and $\eta<0$ (blue) classes. (d) Critical-current noise $\sigma_u$ produced by chemical-potential noise $\sigma_\mu$ at $\mu=100$~meV; the axis starts at $\sigma_u=10^{-6}$, the value used in Figs.~\ref{fig:pos} and \ref{fig:neg}. Exchange $J/\mu=0.2$, barrier $Z=0.5$.}
  \label{fig:am}
\end{figure}

\emph{An altermagnetic platform.}---Altermagnets are compensated collinear magnets with spin-split bands and no net magnetization \cite{Smejkal2022PRX,Smejkal2022,Krempasky2024,Reimers2024}. They are being explored for spin torques \cite{Vakili2025stt,Vakili2026neel} and spin qubits \cite{Vakili2026spinqubit,AbadilloUriel2026}, and Josephson junctions of altermagnetic CrSb have been reported \cite{Esin2026,Gao2026CrSb}. In a superconductor/\allowbreak altermagnet/\allowbreak superconductor junction, tuning through a $0$--$\pi$ transition suppresses the first harmonic and leaves a pronounced second \cite{Ouassou2023,BeenakkerVakhtel2023,Lu2024phi,Sun2025}. We use the continuum model of Ref.~\cite{Lu2024phi}, an altermagnet between two $s$-wave superconducting leads [Fig.~\ref{fig:am}(a)]. The Hamiltonian of the altermagnet is
\begin{equation}
H=\mu\Bigl(\frac{k^2}{k_F^2}-1\Bigr)-\delta\mu\pm\frac{2J}{k_F^2}k_xk_y.
\label{eq:am_h}
\end{equation}
Here $\mathbf k$ is the wave vector, and the two signs are for spin up and down. The leads have chemical potential $\mu$ and Fermi wave vector $k_F$. The gate voltage shifts the chemical potential of the altermagnet by $\delta\mu$. The last term is the altermagnetic exchange splitting, which has strength $J$ and $d$-wave symmetry. The current flows along $x$, so the splitting is largest at $45^\circ$ to the current. The altermagnet has no disorder and is $20/k_F$ long. The leads have a superconducting gap of $0.01\mu$. Each interface has a delta-function barrier of strength $2Z\mu/k_F$. We compute the CPR with Kwant \cite{Groth2014kwant} from a Matsubara Green-function sum with semi-infinite leads. The temperature is $0.025$ of the critical temperature of the leads. We validate the calculation against published results \cite{SM}. The junction energy $U(\delta)$ takes all three shapes of Fig.~\ref{fig:circuit}(a) [Fig.~\ref{fig:am}(b)], and the gate voltage \emph{selects the class} [Fig.~\ref{fig:am}(c)]. About a third of the gate voltages give the $\eta>0$ class, a quarter the $\eta<0$ class and the rest a single well, in interleaved windows a few meV wide. At $\delta\mu/\mu=-0.6$ the harmonic vector is
\begin{equation}
E_m/|E_2|\approx\bigl(0.41,\,1,\,0.06,\,-0.14,\,-0.02,\,0.05\bigr),
\label{eq:am_vector}
\end{equation}
with $\eta\approx2.5$. With this vector on all branches at $|E_2|/E_C=200$, the computational states are degenerate with vanishing $\mathcal N_{\mathcal Q}$, and the next state lies about $|E_2|/3$ above. Figure~\ref{fig:pos} uses the same vector at $|E_2|/E_C=10$. With Rashba coupling $\alpha_R$ the supercurrent depends on the N\'eel-vector direction \cite{Vakili2026neel}. The junction then acquires the odd term only where symmetry allows: for a N\'eel-vector component along the current, or for unequal barriers with an out-of-plane N\'eel vector. At the $\eta<0$ point $\delta\mu/\mu=0.3$, $|A|\simeq20(\alpha_Rk_F/\mu)|E_2|$. The qutrit's odd term ($E_a=0.05$) therefore needs only $\alpha_Rk_F/\mu\approx0.0025$, or $\alpha_R\approx0.15$~meV\,nm at $\mu=100$~meV and the free-electron mass. However, identical junctions with one N\'eel vector acquire the same $A$, which cannot split the qutrit. The intrinsic route needs a N\'eel-vector orientation or Rashba coupling that differs between branches. We have not modeled such a loop, so we use side junctions. For $\eta>0$ the odd term breaks S3, so the N\'eel vector must stay out of plane and the barriers symmetric \cite{SM}.

\emph{Gate-voltage noise.}---The gate voltage that selects the class also moves the harmonics, so its noise $\sigma_\mu$ acts as critical-current noise, $\sigma_u=\lambda\sigma_\mu/\mu$, where $\lambda=d\ln E/d(\delta\mu/\mu)$ and $E$ is the energy that carries the qutrit splitting [Fig.~\ref{fig:am}(d)]. For $\eta>0$ that energy is $\Eodd=\sum_{m\ {\rm odd}}E_m$, half the splitting of the $0$ and $\pi$ wells, and $\lambda\approx400$. Then $\sigma_u=10^{-6}$ needs sub-neV stability of $\mu=100$~meV, and a realistic $0.1~\mu$eV gives a dephasing time of about $50$~ns per branch for $|E_2|/h=10$~GHz. The large slope is structural. The class windows are narrow \emph{because} the first harmonic sweeps fast, their width scaling as $8|E_2|/|dE_1/d(\delta\mu/\mu)|$. The derivative that makes a window a few meV wide therefore makes the device limited by gate-voltage noise. Gate-charge noise already limits gatemons \cite{Larsen2020}. The $\eta<0$ class is less exposed, since its splitting comes from the side junctions and S1 protects the main ones. At the $\eta<0$ point $\lambda\approx60$ for $|E_1|$ and $15$ for $|E_2|$, and $0.1~\mu$eV gives microseconds per branch, or longer when the branches share one gate \cite{SM}.

\emph{Conclusion.}---We have shown that a three-junction loop hosts protected qutrits in both sign classes of the second harmonic, whichever platform supplies the junctions. We estimate the coherence for $|E_2|/h=10$~GHz, $Q=10^6$ and $20$~mK, with the noise amplitudes above and $1\,\mu\Phi_0$ of flux noise in every loop. Relaxation is where both classes gain over a transmon, by two to three orders of magnitude. A transmon qutrit \cite{Koch2007,Peterer2015} modeled with the same $Q$ and $\sigma_u$ has $T_1\approx20$--$30~\mu$s and, from critical-current noise, $T_\varphi\approx30~\mu$s. The $\eta>0$ device has $T_1\gtrsim50$~ms at the worst offset charge (unbounded at $\mathbf n_g=0$), and the $\eta<0$ qutrit has $T_1\approx10$~ms. Dephasing therefore sets the coherence time. The $\eta>0$ device has a Ramsey time $T_2^*\approx5$--$10~\mu$s, set by critical-current noise and quasiparticles. For the $\eta<0$ qutrit, quasiparticles limit $T_2^*$ to $5$--$15~\mu$s without tracking. Parity tracking raises it to about $0.1$~ms, limited by the side junctions' quasiparticle channel and by critical-current noise, with charge noise (seconds) and flux noise (a millisecond) far behind. A Hahn echo refocuses critical-current, charge and flux noise, but not quasiparticle events. The computational levels are $0.5$--$1$~GHz apart, so $10$--$35\%$ of the population is thermally excited at $20$--$40$~mK. Both classes must therefore be initialized by active reset rather than by waiting.

Which class to use depends on what limits the device. The $\eta>0$ class has the exact charge selection rule and the faster gates, under $10$~ns against a few tens of nanoseconds. The $\eta<0$ class is insensitive to charge noise, can be driven directly, and its quasiparticle errors can be corrected. For altermagnetic junctions, gate-voltage noise would dominate the estimates above. Interference-based elements realize the $\eta>0$ class \cite{Gladchenko2009,Bell2014,Smith2020,Larsen2020,Roverch2026,Messelot2026,Zhurbina2026}, and the selection rule is exact when their two arms are balanced. An arm imbalance adds the odd term, which breaks S3, and the flux through the element's own loop tilts the $0$ and $\pi$ wells at first order \cite{SM}. Twisted $d$-wave junctions supply the $\eta<0$ class natively \cite{Sigrist1995,Can2021,Tummuru2022,Zhao2023,Volkov2024}, with a twist angle fixed at fabrication and a bath of nodal quasiparticles. With side junctions they too host the $\eta<0$ qutrit. The remaining priorities are a fast parity detector, the stability of small junctions and gate voltages, and quasiparticle physics beyond the tunnel-junction model for transparent weak links.

\textit{Acknowledgments.}---This work was supported by the National Science Foundation/EPSCoR RII Track-1: Emergent Quantum Materials and Technologies (EQUATE), Award OIA-2044049 and by the UNL Grand Challenges catalyst award entitled Quantum Approaches addressing Global Threats. The calculations and writing of the paper were reviewed by Claude Opus 5.5.

\bibliography{main}

\end{document}